\documentclass[10pt,preprintnumbers,amsmath,twocolumn,amssymb,floatfix]{revtex4}

\usepackage{graphicx, verbatim, subfigure}
\usepackage{bm}

\begin{document}

\title{Phase behaviour of binary mixtures of diamagnetic colloidal
  platelets in an external magnetic field}

\author{Jonathan Phillips} 
\address{H.H. Wills Physics Laboratory, University of Bristol,
  Royal Fort, Tyndall Avenue, Bristol BS8 1TL, United Kingdom}
\author{Matthias Schmidt}
\address{Theoretische Physik II, Physikalisches Institut, 
  Universit\"at Bayreuth, Universit\"atsstra\ss e 30, D-95440 Bayreuth, 
  Germany}
\address{H.H. Wills Physics Laboratory, University of Bristol,
  Royal Fort, Tyndall Avenue, Bristol BS8 1TL, United Kingdom}

\begin{abstract}
  Using fundamental measure density functional theory we investigate
  paranematic-nematic and nematic-nematic phase coexistence in binary
  mixtures of circular platelets with vanishing thicknesses. An
  external magnetic field induces uniaxial alignment and acts on the
  platelets with a strength that is taken to scale with the platelet
  area. At particle diameter ratio $\lambda=1.5$ the system displays
  paranematic-nematic coexistence. For $\lambda=2$, demixing into two
  nematic states with different compositions also occurs, between an
  upper critical point and a paranematic-nematic-nematic triple
  point. Increasing the field strength leads to shrinking of the
  coexistence regions. At high enough field strength a closed loop of
  immiscibility is induced and phase coexistence vanishes at a double
  critical point above which the system is homogeneously nematic. For
  $\lambda=2.5$, besides paranematic-nematic coexistence, there is
  nematic-nematic coexistence which persists and hence does not end in
  a critical point. The partial orientational order parameters along
  the binodals vary strongly with composition and connect smoothly for
  each species when closed loops of immiscibility are present in the
  corresponding phase diagram.
\end{abstract}


\date{23 December 2010, to appear in J. Phys.: Condensed Matter}
\maketitle

\section{Introduction}
Dispersions of colloidal plateletlike particles, such as gibbsite
\cite{vanderbeek04, vanderbeek05}, montmorillonite
\cite{pizzey.c:2004.a, connolly.j:2007.a} or iron-rich beidellite
\cite{michot.lj2006.a, paineau.e:2009.a}, are susceptible to the
influence of magnetic fields, since the particles possess nonvanishing
diamagnetic anisotropy. When a magnetic field is applied to an
initially isotropic (I) platelet dispersion, the field induces
orientational order in the system, thus breaking the rotational
symmetry; an orientationally ordered paranematic (P) phase
results. The paranematic phase has interesting optical properties,
similar to those of the nematic (N) phase. When observed through
crossed polarisers, samples of gibbsite suspensions have been shown to
exhibit field-induced birefringence \cite{vanderbeek.d:2006.a}.
Birefringence gradients have also been theoretically modelled for a
simple model system \cite{reich.h:2010.a}.  The effects of a magnetic
field on montmorillonite platelets were studied in
Ref.~\cite{connolly.j:2007.a} and on hematite platelets in
Ref.~\cite{ozaki.m:1990.a}. Unlike the gibbsite platelets, hematite
platelets are ferromagnetic and an I-N transition was not observed;
rather the authors found a clustering effect whereby chains of
particles were formed due to the platelet-platelet interactions. Such
clustering has also been observed in simulations
\cite{satoh.a:2009.a}. Experimental investigations of gibbsite
platelets, whereby the suspensions were exposed to magnetic fields
\cite{vanderbeek.d:2006.a, vanderbeek.d:2008.a}, showed that a
paranematic phase occurs in these systems.

The phase behaviour of rods in external aligning fields is
well-studied, see e.g.\ \cite{khokhlov.ar:1982.a,
  palffymuhoray.p:1982.a, fraden.s:1993.a, varga.s:1998.a,
  tang.jx:1993.a, graf.h:1999.a} for studies of one-component
systems. In Ref.~\cite{khokhlov.ar:1982.a} the effect of external
fields on the phase behaviour of rigid rods, freely jointed rods and
semiflexible rods was analysed with Onsager theory
\cite{onsager.l:1949.a}: the P-N transition was found to terminate at
a critical point in all three cases at a high enough field
strength. In Ref.~\cite{palffymuhoray.p:1982.a} the theories of Landau
and de Gennes \cite{degennes} and of Maier and Saupe
\cite{maier.w:1959.a} were used to analyse the effects of an applied
field in the nematic phase.  The magnetic field-induced birefringence
in solutions of tobacco mosaic virus (TMV) particles was studied in
Ref.~\cite{fraden.s:1993.a} both experimentally and theoretically
(using extensions of Onsager theory). In Ref.~\cite{varga.s:1998.a}
the phase behaviour of monodispese rods with varying aspect ratio was
studied using the Parsons-Lee scaling \cite{parsons.jd:1979.a,
  lee.sd:1987.a} of the Onsager functional. It was found that the
bifurcation density decreases with increasing field strength.  The
nematic order of model goethite nanorods in a magnetic field was
investigated in Ref.~\cite{wensink_goethite}, also using Parsons-Lee
theory. The goethite rods were modelled as charged spherocylinders
with a permanent magnetic moment along the long axis of the rods. This
encourages the rods to align parallel to the field at low field
strengths. However, goethite rods possess a negative diamagnetic
susceptibility which leads to alignment perpendicular to the field at
higher field strengths. These competing effects were found to yield
rich phase diagrams including biaxial arrangements of the
particles. The phase separation in suspensions of semiflexible
\textit{fd}-virus particles was studied in Ref.~\cite{tang.jx:1993.a},
where a P-N phase transition was found. Effects of an external field
on the isotropic, nematic and smectic-A phases of spherocylinders were
compared with simulations and theory in Ref.~\cite{graf.h:1999.a},
with results from both approaches being in good agreement.

Even when neglecting positionally ordered phases (such as columnar and
crystal phases), the bulk phase behaviour of binary mixtures of
non-spherical colloidal particles can be very rich, often including
isotropic-isotropic (I-I), isotropic-nematic (I-N) and nematic-nematic
(N-N) phase coexistence, depending on the value of the size asymmetry
parameter of the two species. The asymmetry parameter may quantify the
difference in thickness or lengths of the two species. An example are
binary mixtures of thick and thin hard rods in an external
field~\cite{varga.s.:2000.a, matsuda.h:2004.a, dobra.s:2006.a}. A
general feature of the phase behaviour of binary mixtures is a
widening of the biphasic region on increasing the asymmetry
parameter. In a certain range of the size asymmetry there is typically
an I-N-N and/or an I-I-N triple point. Coexistence between two nematic
states may or may not end in a critical point depending on the system
under study and the value of the asymmetry parameter. A well-studied
system is the Zwanzig model for binary hard platelets, where the
particles are restricted to occupy three mutually perpendicular
directions. This was shown to exhibit rich bulk phase diagrams
\cite{harnau02,bier04,harnau08}.
We recently explored the phase behaviour of binary mixtures of hard
platelets with zero thickness and continuous orientations
\cite{phillips.j:2010.a} using fundamental measure theory (FMT).

Platelets can be characterized by a diamagnetic susceptibility tensor
that is diagonal in the platelet frame of reference, with components
$\chi_\parallel$ in the platelet plane and $\chi_\perp$ normal to it.
The diamagnetic anisotropy $\Delta \chi \equiv
\chi_\parallel-\chi_\perp\neq 0$, in general, and it may be positive
or negative depending on the properties of the platelet material. For
Gibbsite platelets $\Delta \chi<0$, therefore the platelets tend to
align with their normals perpendicular to the direction of the applied
field. In order for the platelets to align uniaxially in the presence
of the field, the samples were placed on a central stage and rotated
in a horizontally applied field. In Ref.~\cite{reich.h:2010.a}, FMT
was used to study the effects of an external field on the phase
behaviour of monodisperse platelets.  It was found that above a
critical field strength the P-N coexistence ceases to exist and the
system is homogeneously nematic. In Ref.~\cite{vandenpol.dme:2008.a},
van~den~Pol \textit{et. al.} have experimentally investigated the
general phase behaviour of the boardlike goethite colloidal particles
($\alpha$-FeOOH) in the presence of an external magnetic field. The
particles were found to align parallel to a small magnetic field and
perpendicular to a large magnetic field; this had already been known
since the observations of Lemaire \textit{et. al.}
\cite{lemaire.bj:2002.a}. This effect is due to the particles having a
permanent magnetic moment along their long axis but the magnetic easy
axis being the short axis. An exciting prospect is that suspensions of
beidellite platelets, which have a disk-like morphology, have recently
been shown to undergo an I-N transition \cite{michot.lj2006.a} and the
nematic phase aligns strongly in the presence of an externally applied
magnetic or electric field \cite{paineau.e:2009.a}. These platelets
possess a \textit{positive} diamagnetic susceptibility and, as such, a
simpler experimental setup would be required to investigate the P-N
transition; the platelets are expected to align with their normals
parallel to the magnetic field.

Since Rosenfeld's pioneering work
\cite{rosenfeld89,rosenfeld94convex,rosenfeld95convex} there has been
much interest in the development of FMT for non-spherical particles,
see e.g.\ \cite{hansen-goos}. In the current investigation we use the
FMT of Ref.~\cite{phillips.j:2010.a}, which is the mixtures
generalization of the theory proposed in Ref.~\cite{esztermann06}, to
study binary mixtures of diamagnetic platelets in a magnetic field. We
consider three different size ratios representative of the different
topologies of the bulk phase diagram and the full range of external
field strengths. We investigate how the phase behaviour for each of
these three size ratios changes on increasing the external field
strength, which we take to scale with the platelet area and to induce
uniaxial alignment.  

This paper is organised as follows. In Sec.~\ref{sec:theory} we
outline the density functional theory for the model system. The phase
diagrams and results for order parameters are presented in
Sec.~\ref{sec:results} and we conclude in Sec.~\ref{sec:conclusions}.
\section{Theory}
\label{sec:theory}
\subsection{Pair Interactions, Model Parameters and External Orienting Field}
\label{model}
We consider a binary mixture of hard circular platelets with vanishing
thickness and continuous positional and orientational degrees of
freedom. Particles of species 1 and 2 possess radii $R_1$ and $R_2$,
respectively, and we take $R_2>R_1$. The pair potential $u_{ij}$
between two particles of species $i$ and $j$, where $i,j=1,2$, models
hard core exclusion and is hence given by
\begin{equation}
  {u_{ij}(\bf{r}-\bf{r}', \boldsymbol{\omega},\boldsymbol{\omega}')=
    \begin{cases}\infty&\text{if particles overlap}\\
      0&\text{otherwise, } \end{cases}
}
  \label{eq:potential}
\end{equation}
where $\textbf{r}$ and $\textbf{r}'$ are the positions of the particle
centres and $\boldsymbol{\omega}$ and $\boldsymbol{\omega}'$ are unit
vectors indicating the particle orientations (normal to the particle
surface). As a control parameter that characterises the radial
bidispersity we use the size ratio
\begin{equation}
\label{eq:size_ratio}
\lambda=\frac{R_2}{R_1} > 1.
\end{equation}
The effect of a magnetic field on the diamagnetic platelets is
described by an external potential for each species,
\begin{equation}
  V_\textrm{ext}^{(i)}(\theta)= \beta^{-1}W_i\sin^2\theta,\hspace{5mm}i=1,2,
  \label{eq:external_potential}
\end{equation}
where $\beta=1/(k_BT)$, with $k_B$ being the Boltzmann constant and
$T$ absolute temperature; $\theta$ is the angle between the platelet
orientation $\boldsymbol{\omega}$ and the direction of the external
field. The strength of the external potential of species $i$ is
related to the material and field properties via
\begin{equation}
  W_i=-\frac{\beta}{4}B^2\Delta\chi_i
\end{equation}
where $B$ is the magnetic flux density (measured in $\textrm{T})$ and
$\Delta \chi_i=\chi_\parallel^{(i)}-\chi_\perp^{(i)}$ is the
diamagnetic susceptibility anisotropy (with units of JT$^{-2}$) of
species $i$, with $\chi_\parallel^{(i)}$ and $\chi_\perp^{(i)}$ being
the susceptibilities perpendicular and parallel to the field,
respectively \footnote{In the case of an electric field,
  $W_i=-E^2\Delta \epsilon_i/2$, where $E$ is the electric field
  strength measured in Vm$^{-1}$. $\Delta \epsilon_i
  =\epsilon_\parallel^{(i)}-\epsilon_\perp^{(i)}$ is the dielectric
  anisotropy. Beidellite platelets possess a negative dielectric
  anisotropy.}.  In general both $W_1$ and $W_2$ constitute further
control parameters. We restrict ourselves in the following to special
cases and assume that $W_i$ scales with the platelet area, i.e.\ $W_i
\sim R_i^2$. This implies the relationship $W_2=\lambda^2 W_1$, and we
hence take $W_1$ to be our second control parameter, besides the size
ratio $\lambda$ itself. Scaling with the platelet area is motivated by
the assumption that the platelets interact with an external field in a
manner proportional to their mass (neglecting any effects of
thickness). We could well envisage that scaling the strength of the
potential e.g.\ with the radius would be another, different yet
sensible, choice. We neglect platelet-platelet interactions due to
induced dipoles because of their small magnitude, see e.g.\ the
discussion in Ref.~\cite{reich.h:2010.a}.

The thermodynamic state is characterised by two dimensionless
densities $c_1 = \rho_1 R_1^3$ and $c_2 = \rho_2 R_1^3$, where
$\rho_1$ and $\rho_2$ are the number densities of the two species,
$\rho_i=N_i/V$, where $N_i$ is the number of particles of species
$i=1,2$ and $V$ is the system volume.  The composition (mole fraction)
of the (larger) species 2 is $x=\rho_2/(\rho_1+\rho_2)$ and the total
dimensionless concentration is $c=(\rho_1+\rho_2)R_1^3=c_1+c_2$.

\subsection{Density Functional Theory}
Density functional theory (DFT) is formulated on the level of the
one-body density distributions $\rho_i(\bf{r},\boldsymbol{\omega})$ of
each species $i$. The variational principle \cite{evans.r:1979.a}
asserts that minimising the grand potential functional $\Omega$ yields
the true equilibrium density profile,
\begin{equation}
  \frac{\delta\Omega([\rho_1,\rho_2],\mu_1,\mu_2,V,T)}
       {\delta \rho_i(\textbf{r},\boldsymbol{\omega})}=0,\hspace{5mm}i=1,2
\label{eq:minimization_principle}
\end{equation}
where $\mu_i$ is the chemical potential of species $i$. The grand
potential functional is given by
\begin{align}
  \Omega([\rho_1,&\rho_2],\mu_1,\mu_2,V,T)= 
  F_\textrm{id}([\rho_1,\rho_2],V,T)\notag \\
  & + F_\textrm{exc}([\rho_1,\rho_2],V,T) 
  + \sum_{i=1}^2\int d \textbf{r} \int d \boldsymbol{\omega} 
  \rho_i (V_\textrm{ext}^{(i)}(\textbf{r},\boldsymbol{\omega})-\mu_i),
  \label{eq:grand_potential}
\end{align}
where the spatial integral (over $\textbf{r}$) is over the system
volume $V$ and the angular integral (over $\boldsymbol{\omega}$) is
over the unit sphere. The inter-particle interactions are described by
the excess (over ideal gas) contribution to the Helmholtz free energy,
$F_\textrm{exc}([\rho_1,\rho_2],V,T)$. We skip the explicit definition
of the FMT approximation here; this can be found in
Ref.\ \cite{phillips.j:2010.a}. The free energy functional for a
binary ideal gas of uniaxial rotators is given by
\begin{align}
  \beta F_\textrm{id}([\rho_1,\rho_2],V,T) = 
  \sum_{i=1}^2&\int{d\textbf{r}}\int d\boldsymbol{\omega}
  \rho_i(\textbf{r},\boldsymbol{\omega})\notag \\ 
  &\times[\ln(\rho_i(\textbf{r},\boldsymbol{\omega})\Lambda_i^3)-1],   
\label{eq:ideal}
\end{align}
where $\Lambda_i$ is the (irrelevant) thermal wavelength of species
$i$.

For bulk fluid states (i.e.\ with the density distribution not
depending on $\textbf{r}$) the orientational distribution functions
(ODFs), $\Psi_i(\theta)$, $i=1,2$, are related to the one-body density
distributions by
$\rho_i(\textbf{r},\boldsymbol{\omega})=\rho_i\Psi_i(\theta)$. There
is no dependence of the ODF on the azimuthal angle $\phi$ since the
platelets are uniaxial rotators, and we assume that only uniaxial
states are formed.  A powerful feature of DFT is that
$V_\textrm{ext}^{(i)}(\textbf{r},\boldsymbol{\omega})$
(\ref{eq:external_potential}) appears explicitly in the grand
potential and therefore enters straightforwardly into the minimisation
procedure (\ref{eq:minimization_principle}); see the appendix for the
explicit form of the corresponding Euler-Lagrange equations that we
solve numerically.

The requirements for phase coexistence between two phases \textit{A}
and \textit{B} are the mechanical and chemical equilibria between the
two phases and the equality of temperature in the two coexisting
phases (which is trivial in hard-body systems). Hence we have the
non-trivial conditions: the equality of pressure $p^A=p^B$ and the
equality of chemical potentials $\mu_i^A=\mu_i^B$, where $i=1,2$ again
labels the species, and $A,B$ labels the phase.  We calculate the
total Helmholtz free energy $F=F_\textrm{id}+F_\textrm{exc}$
numerically by inserting $\Psi_i(\theta)$ into the free energy
functional. Likewise, the pressure can be obtained numerically as $p =
-F/V+\sum_{i=1}^2\rho_i\partial(F/V)/ \partial \rho_i$ and the
chemical potentials as $\mu_i= \partial (F/V) / \partial \rho_i$.  We
define a reduced pressure $p^*=\beta p R_1^3$ and reduced chemical
potentials $\mu_i^*=\beta \mu_i$.  The equations for phase coexistence
are three equations for four unknowns (two statepoints each
characterised by two densities) hence regions of two-phase coexistence
depend parametrically on one free parameter (which can be chosen
arbitrarily, e.g. as the value of composition $x$ in one of the
phases) and are solved numerically with a Newton-Raphson procedure
\cite{press.wh:2007.a}. The resulting set of solutions yields the
binodal. P-N-N triple points are located where the P-N and N-N
coexistence curves cross. In the fieldless case there is, of course,
not a paranematic phase, but an isotropic phase.

We characterize orientationally ordered phases (P and N) of the binary
mixture by two partial order parameters, $S_1$ and $S_2$, defined by
\begin{equation}
\label{eq:order_parameter}
S_i = 4\pi \int_0^{\pi/2}d\theta\sin (\theta) \Psi_i(\theta)P_2(\cos\theta),
\end{equation}
where $P_2(\cos\theta)=(3\cos^2\theta-1)/2$ is the second
Legendre polynomial in $\cos \theta$.

\section{Results}
\label{sec:results}
\begin{figure}
  \centering 
  \subfigure{\includegraphics[width=0.45\textwidth,clip]{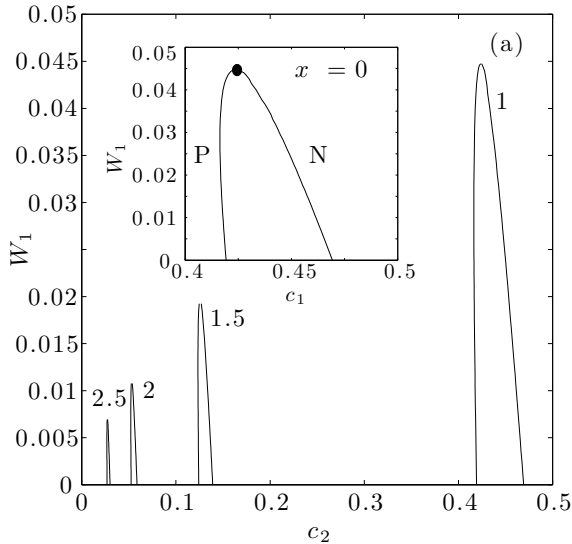}}
  \subfigure{\includegraphics[width=0.45\textwidth,clip]{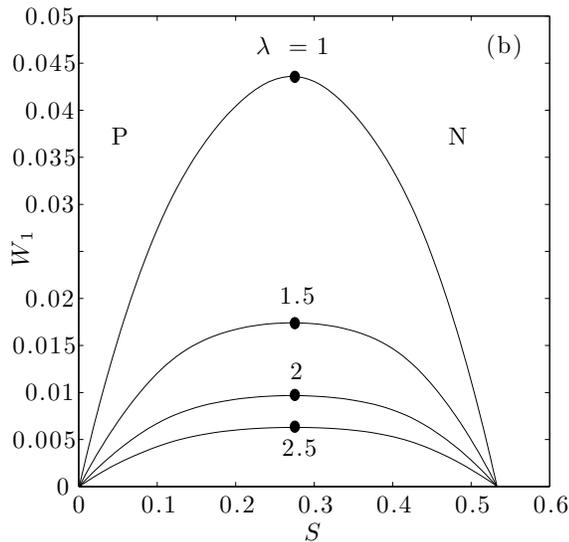}}
  \caption{Paranematic-nematic phase diagram of the one-component
    system(s).  (a) Behaviour of the pure system of species 2
    (i.e.\ $x=1$). The variation of the paranematic (P) and nematic
    (N) coexistence concentrations, $c_2$ (horizontal axis) with the
    strength of the aligning field $W_1$ (vertical axis) is shown for
    $\lambda=1, 1.5, 2$ and $2.5$ (from right to left, as
    indicated). The inset shows the phase diagram for the pure system
    of species 1 (i.e.\ $x=0$). This is equivalent to the case
    $\lambda=1$ in the main plot when identifying the horizontal
    axes. The critical point is depicted as a filled circle. (b)
    Variation of the orientational order parameter $S=S_2$ along the
    paranematic and nematic branches of the binodal (horizontal axis)
    of the pure system of species 2, with increasing field strength
    $W_1$ (vertical axis) for the same size ratios as in (a). Critical
    points are depicted as filled circles.}
  \label{figure1} 
\end{figure}

We first review the behaviour of the pure system under the influence
of an aligning field~\cite{reich.h:2010.a}.  The inset of
Fig.~\ref{figure1}(a) shows the phase diagram for a system composed of
particles of species~1 only. Upon increasing the field strength $W_1$,
the coexisting concentrations $c_1$ initially shift to lower
values. The biphasic density gap decreases slightly as the strength of
the external potential is increased. At approximately $W_1=0.02$, the
paranematic coexistence concentration starts to increase, while the
nematic coexistence concentration continues to decrease. The two
branches of the binodal meet at a critical point at
$c_1^{\textrm{crit}}=0.42$. For $W_1>W_1^\textrm{crit}=0.045$, there
is no longer a phase transition and complete destabilisation of the
P-N transition results in a homogeneous nematic phase. For the case of
equal sizes of the two components, the pure system of species 2
possesses the same phase diagram, see the binodal for the case
$\lambda=1$ in main plot of Fig.~\ref{figure1}(a). However, due to the
definition of $c_2$ (recall that $c_2=\rho_2 R_1^3$, using the radius
of species 1 in order to obtain a dimensionless quantity) and the
scaling of $W_2$ with the square of the size ratio ($W_2 = \lambda^2
W_1$), the phase diagram of the pure system of species 2 displays
strong variation with size ratio $\lambda$, as shown in
Fig.~\ref{figure1}(a). A shift to both smaller values of $c_2$ and of
$W_1$ occurs upon increasing the value of $\lambda$. However, this
effect is entirely due to the choice of coordinates, which for both
pure systems are related via $c_2=c_1/\lambda^3$ and
$W_2=W_1/\lambda^2$.  Numerical values for the location of the
critical point are summarised in Tab.~\ref{critical}.

\begin{table}[ht]
  \begin{center}
  \begin{tabular}{|c|c|c|c|c|c|}\hline
  $\lambda$& $\lambda^2$& $\lambda^3$& $c_2^\textrm{crit}$& $W_1^\textrm{crit}$\\
  \hline
  1   & 1    & 1      & 0.42   & 0.045  \\
  1.5 & 2.25 & 3.375  & 0.124  & 0.019  \\
  2   & 4    & 8      & 0.053  & 0.011  \\
  2.5 & 6.25 & 15.625 & 0.027  & 0.0072 \\
  \hline
\end{tabular}
\end{center}
\caption{Scaling of the location of the critical point with size
  ratio: Critical concentration
  $c_2^{\textrm{crit}}=\lambda^{-3}c^{\textrm{crit}}$ and critical
  field strength $W_1^\textrm{crit}=\lambda^{-2}W^\textrm{crit}$ for a
  range of size ratios $\lambda$, where $c^\textrm{crit}$ and
  $W^\textrm{crit}$ are the critical concentration and field strength
  in the pure system (without species index).}
\label{critical}
\end{table}

The variation of the order parameter $S$ for monodisperse platelets
with field strength is displayed in Fig.~\ref{figure1}(b). In the
field-free case, $W_1=0$, the nematic phase at coexistence possesses
an unusually small order parameter, see e.g.\ the discussion in
Ref.\ \cite{reich07}. For all size ratios, as $W_1$ is increased, the
coexistence value of $S$ in the paranematic phase increases
monotonically, and the value of $S$ along the nematic branch of the
binodal decreases with increasing field strength. This is consistent
with the fact that the coexistence density decreases as the field
strength increases, overcompensating for the ordering effect caused by
the applied field.  At the critical point the nematic order parameter
takes on the value $S_2=0.27$. For increasing values of
$\lambda=1,1.5,2$ and $2.5$, the critical point shifts to smaller
values of field strengths.

\begin{figure}
  \centering 
  \subfigure{\includegraphics[width=0.22\textwidth,clip]{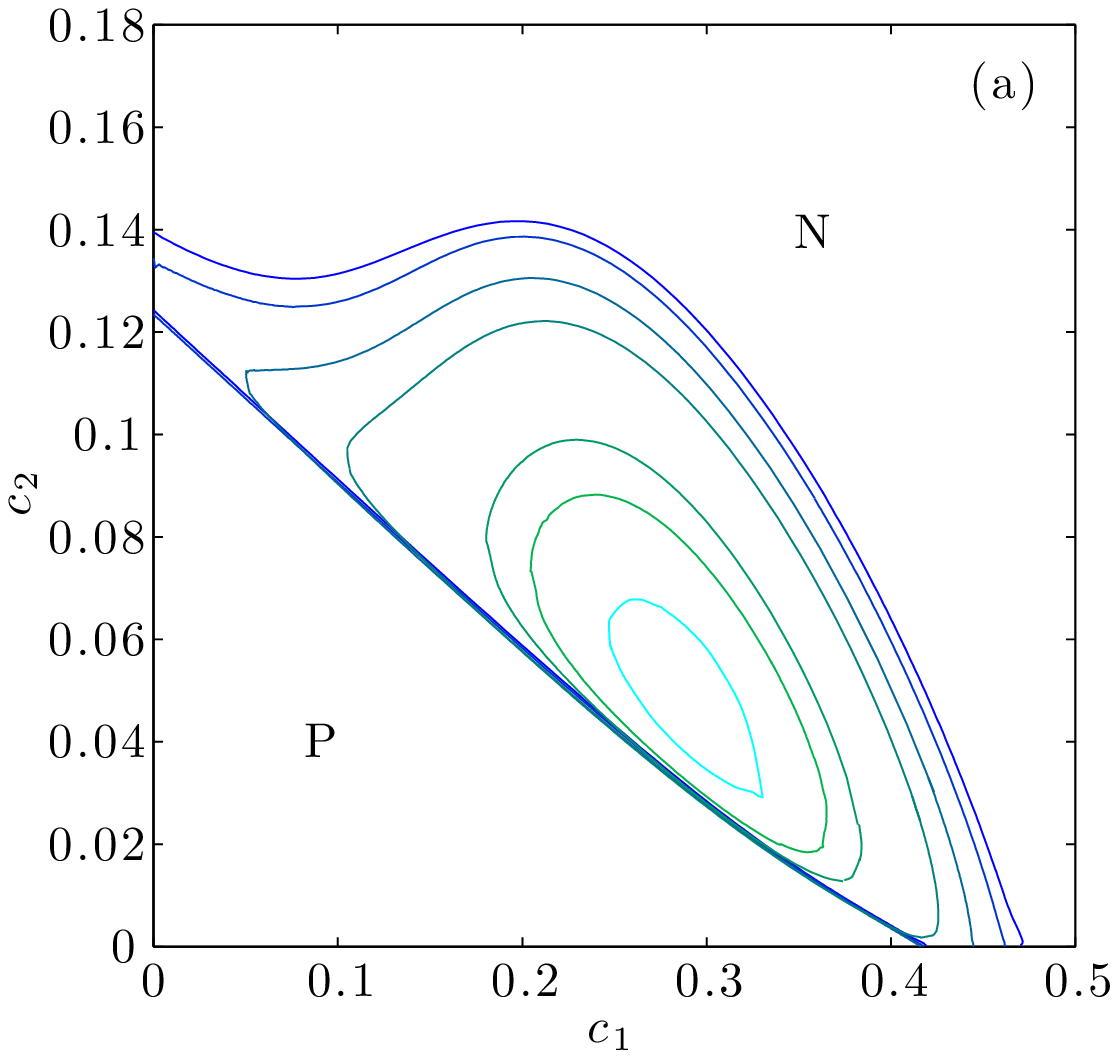}}
  \subfigure{\includegraphics[width=0.22\textwidth,clip]{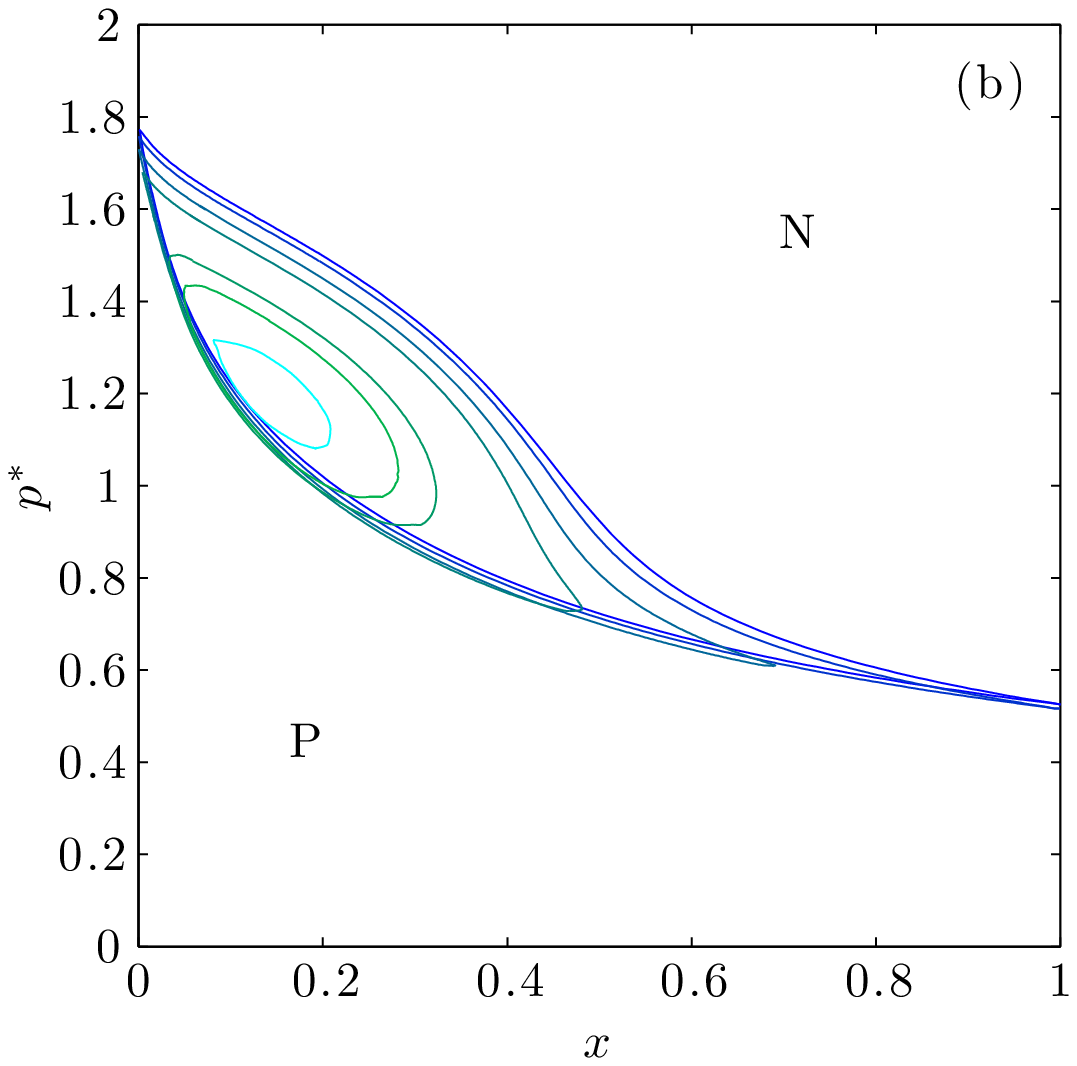}}
  \subfigure{\includegraphics[width=0.22\textwidth,clip]{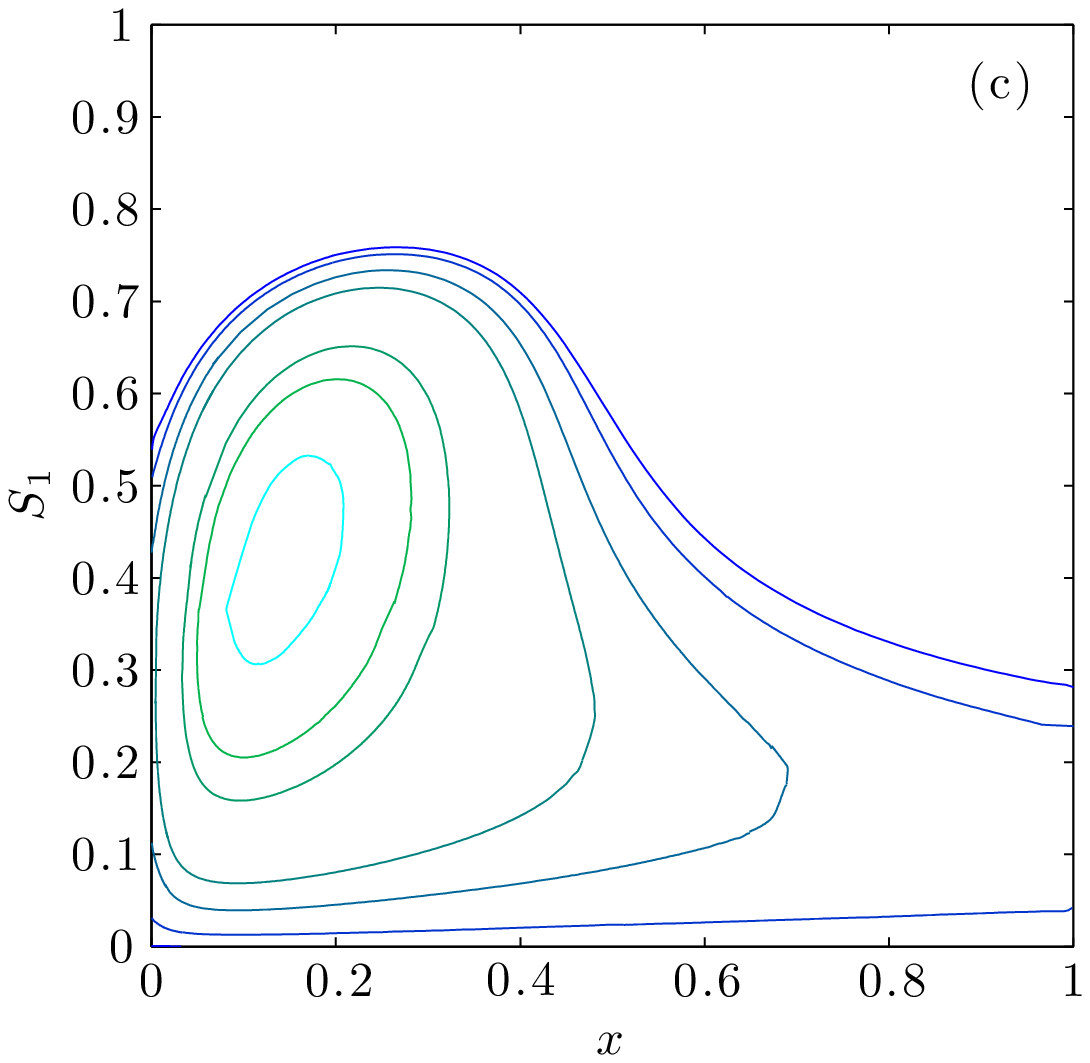}}
  \subfigure{\includegraphics[width=0.22\textwidth,clip]{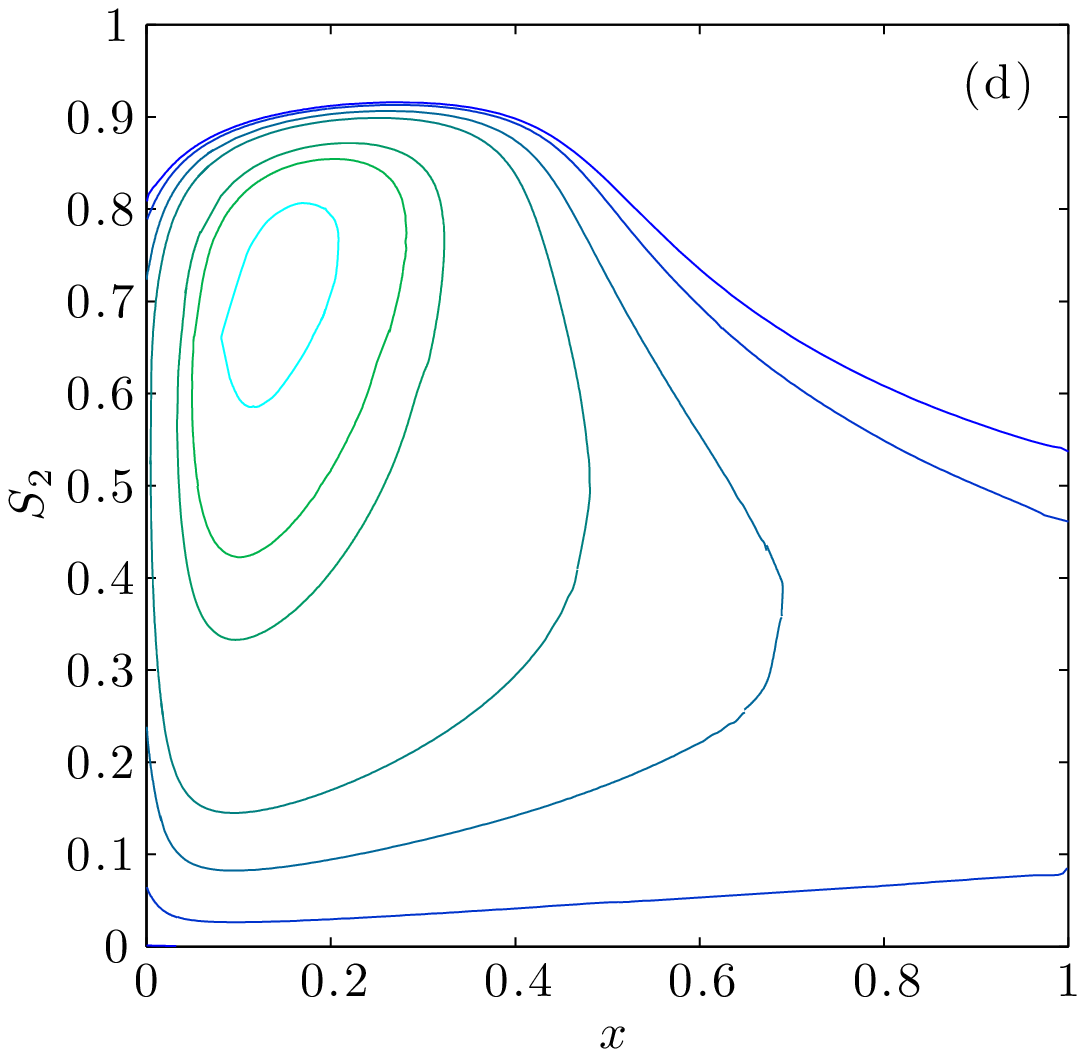}}
  \caption{Phase behaviour of binary platelet mixtures with size ratio
    $\lambda=1.5$ and external potential strength
    $W_1=0,0.01,0.03,0.05,0.07,0.1,0.12$ and $0.15$ (with
    $W_2=\lambda^2 W_1$) from outside to inside. Shown are phase
    diagrams in (a) the $(c_1,c_2)$ representation and (b) in the
    $(x,p^*)$ representation. The partial order parameters $S_i$ along
    the binodal are shown in (c) for species 1 and in (d) for species
    2. }
  \label{figure2} 
\end{figure}

We next consider the binary mixture in the external field and hence
explore the full range of compositions, $0\leq x \leq 1$.  In
Fig.~\ref{figure2} the phase diagram for $\lambda=1.5$ is shown. We
consider a range of external field strengths up to $W_1=0.15$ (which
corresponds to $W_2=\lambda^2 W_1=0.3375$). For the fieldless case,
$W_1=0$, there is I-N phase coexistence over the entire range of
compositions $x$. We display this phase diagram (and subsequent ones)
both in the $(c_1,c_2)$ representation [Fig.~\ref{figure2}(a)] as well
as in the $(x,p^*)$ representation [Fig.~\ref{figure2}(b)]. Tie-lines
are omitted for clarity; in the $(c_1,c_2)$ representation these
connect the lower branch of the binodal to the upper branch in such a
way that the isotropic (or paranematic) phase is rich in (the smaller)
species 1 and the nematic phase is rich in (the larger) species 2. In
the $(x,p^*)$ representation [Fig.~\ref{figure2}(b)] the tie lines are
(trivially) horizontal due to the condition of equal pressures in the
coexisting phases. For $W_1=0.01$, the binodal still connects to the
axes (which correspond to the pure systems). Recall that the P-N
transition still occurs in the pure systems at this field strength,
\textit{cf.}  Fig.~\ref{figure1}(a). However, the isotropic phase has
now become a weakly-ordered paranematic phase. Hence there is P-N
phase coexistence over the entire range of compositions.  The
isotropic phase has been replaced by a paranematic phase, because the
order parameter along the lower branch of the binodal is non-zero, see
Fig.~\ref{figure2}(c) and (d), where the partial nematic order
parameters are shown for species 1 and 2, respectively. On increasing
$W_1$ to $0.03$, the upper and lower branches of the binodal still
persist to the pure system of smaller platelets, consistent with the
findings of Ref.\cite{reich.h:2010.a}. However, the binodal does not
touch the $c_2$-axis, indicating that there is no longer a phase
transition in the pure system of species 2 (we found the critical
field strength for the monodisperse system at $\lambda=1.5$ to be
$0.019$, which is less than $0.03$, Fig.~(\ref{figure1}a)). Hence the
two branches of the binodal connect at a (lower, in pressure) critical
point. Therefore the state of the system changes continuously from
paranematic to nematic for compositions greater than about 0.7 by
increasing the pressure. For compositions less than this value,
increasing the pressure from below the lower branch of the binodal to
the upper branch of the binodal, the system, as before, passes through
a biphasic region. For $W_1=0.05$ the departure of the binodal from
$c_1=0$ (and $x=0$) occurs as is consistent with the critical field
strength being $W_1^\textrm{crit}=0.045$ in the pure system.  The
result is a phase diagram in which the two branches of the binodal
have joined to form a closed loop of immiscibility. There is a larger
range of compositions towards the $x=1$ side of the phase diagram
(approximately $x>0.45$) than towards the $x=0$ side of the phase
diagram (approximately $x<0.02$), where an increase in pressure leads
to a continuous change in state from a paranematic state to a nematic
state. The order parameter of species 1, measured along the binodal
varies with composition [Fig.~(\ref{figure2}c)] such that for $W_1=0,
0.01$ and $0.03$ the paranematic and nematic branches of the binodal
do not connect on the low-composition side of the order parameter
graph since these values of $W_1$ are less than
$W_1^\textrm{crit}=0.045$. For $0.05 \leq W_1 \leq 0.15$ the two
branches of the binodal connect on the low-composition side of the
graph. For $W_1=0$ and $0.01$ the two branches of the binodal do not
connect on the high-composition side of the graph since these values
of $W_1$ are smaller than $W_2^\textrm{crit}=\lambda^2
W_1^\textrm{crit}=0.019$. For $W_1 \geq 0.05$ the partial order
parameters measured along the binodal form closed loops. These islands
become smaller with increasing field strength and eventually coalesce
to a point when the double critical point is reached. The partial
order parameters of species 2 [Fig.~(\ref{figure2}d)] follow a similar
pattern except that the order at a given statepoint is higher than
that for species~1, as one could expect, given that species~2 is of
the larger size.

\begin{figure}
  \centering 
  \subfigure{\includegraphics[width=0.22\textwidth,clip]{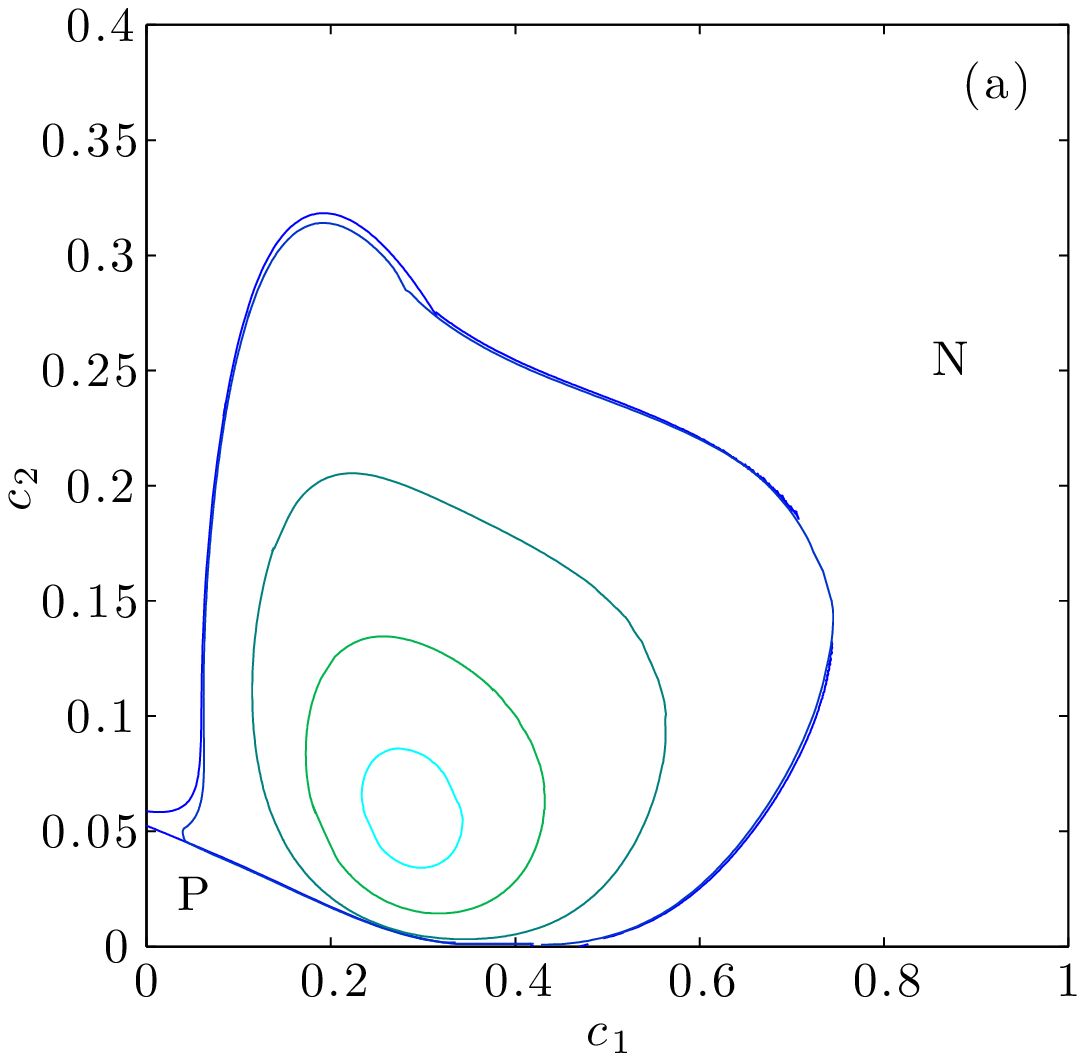}}
  \subfigure{\includegraphics[width=0.22\textwidth,clip]{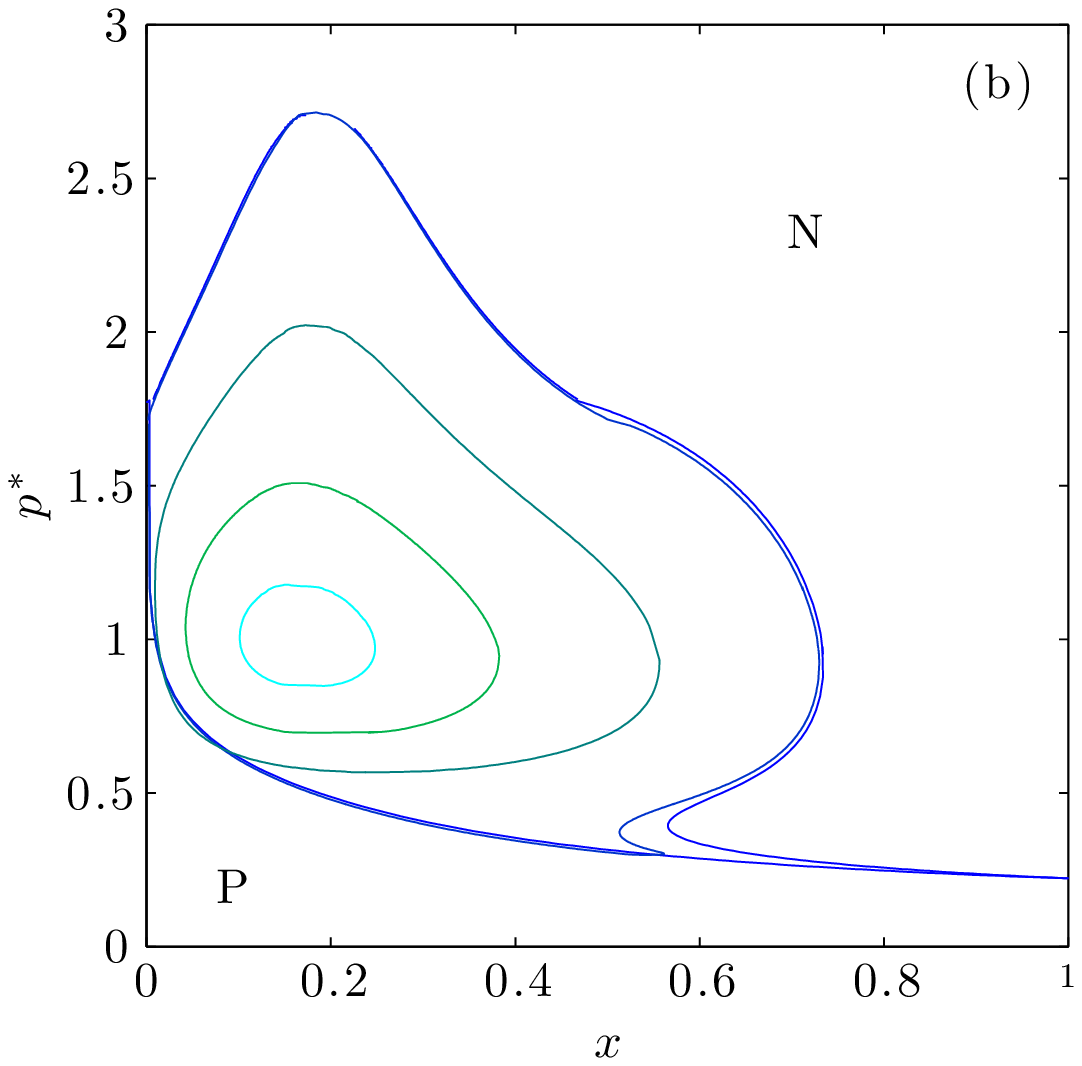}}
  \subfigure{\includegraphics[width=0.22\textwidth,clip]{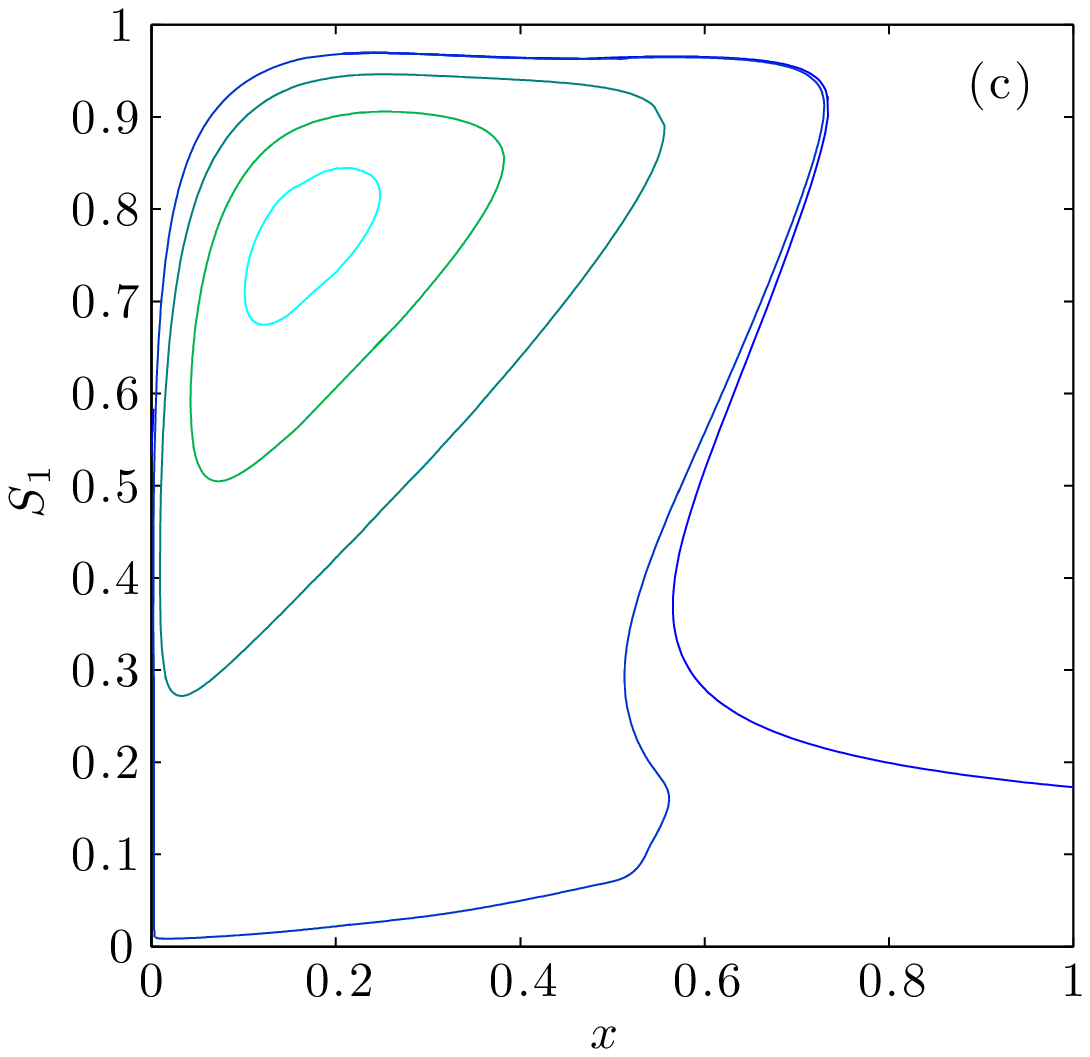}}
  \subfigure{\includegraphics[width=0.22\textwidth,clip]{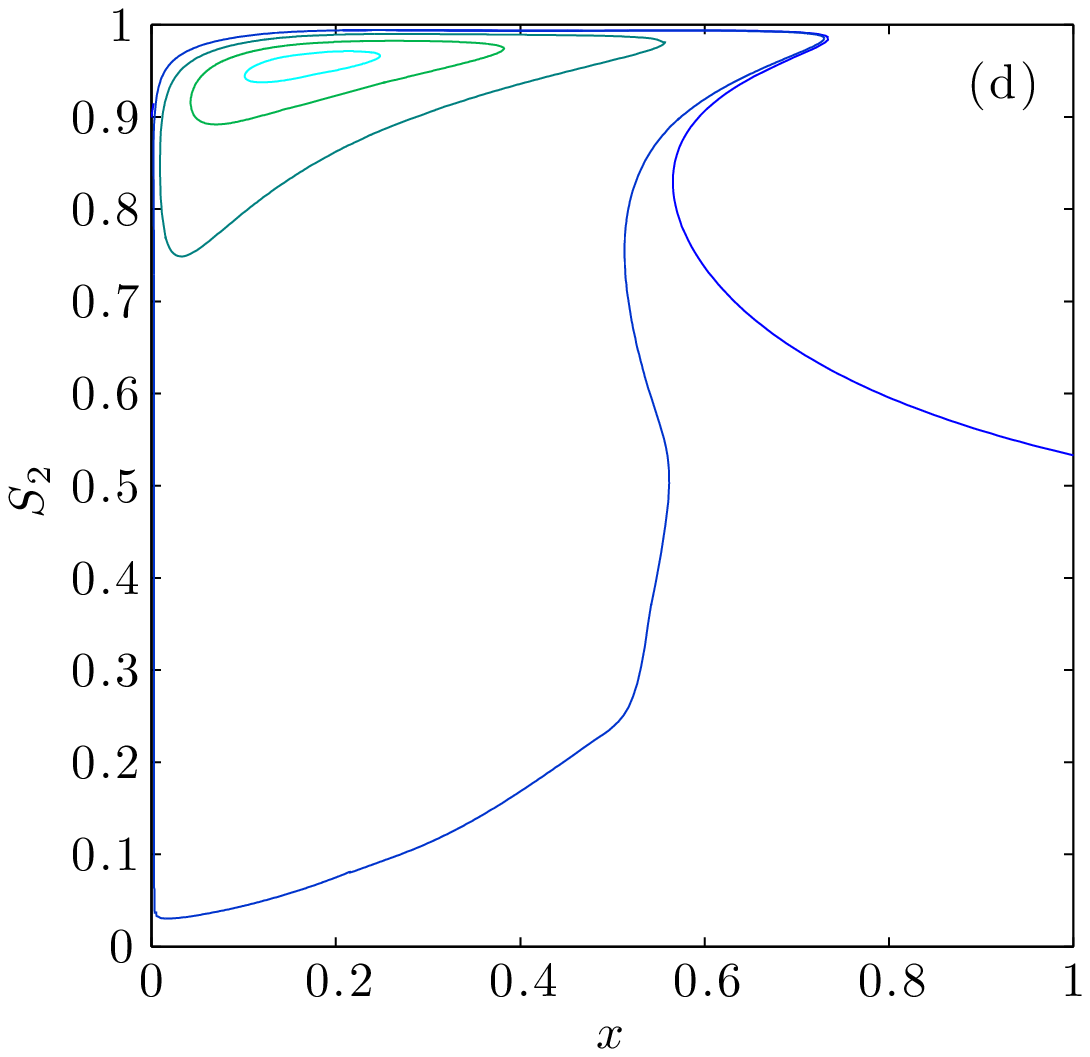}}
  \caption{Same as Fig.~\ref{figure2} except for size ratio
    $\lambda=2$ and for external field strengths $W_1=0,0.02,0.6,1$
    and $1.2$.}
  \label{figure3} 
\end{figure}

In Fig.~\ref{figure3} we show results for $\lambda=2$. Increasing the
size ratio to this value leads to an increase of the size of the I-N
biphasic region \cite{phillips.j:2010.a}. The fieldless case possesses
a reentrant phenomenon whereby the system undergoes the following
change of state when increasing the pressure at fixed mole fraction at
around $x=0.7$ starting in the isotropic region:
I$\rightarrow$I-N$\rightarrow$N$\rightarrow$I-N$\rightarrow$N. In
addition, there is N-N coexistence between a nematic phase rich in
species 1 (N$_1$) and a nematic phase rich in species 2 (N$_2$) ending
in an upper critical point and an I-N-N triple point. Applying a small
field strength of $W_1=0.02$, the binodal no longer reaches the pure
system of species 2, as $W_1=0.02 > W_1^\textrm{crit}=0.011$, which is
the critical field strength for the monodisperse system at
$\lambda=2$. An effect of this is that the reentrant part of the phase
diagram alters: the range of compositions for which the system
undergoes
P$\rightarrow$P-N$\rightarrow$N$\rightarrow$P-N$\rightarrow$N at just
over $x=0.5$ is much smaller than in the fieldless case. The binodal
ends in a tail-like feature at just under $x=0.6$. Aside from these
differences, the rest of the phase boundaries follow closely those of
the fieldless case, though remaining slightly inside those of the
latter. There is N-N coexistence ending in an upper critical
point. The triple point is retained as a P-N-N line in the $(x,p^*)$
representation and a triangle in the $(c_1,c_2)$ representation
although we do not show these features in the plots for clarity.  Upon
increasing the field, triple phase coexistence vanishes, i.e.\ the
triple point collapses onto two-phase coexistence. We have not
calculated the precise value of the external field where this
happens. We expect this value to be different from the values where
the binodal detaches from either of the density axes (i.e.\ differ
from the critical field strengths in the pure systems).  Applying a
field strength $W_1=0.6$, which is much greater than the critical
field strength for the monodisperse case, $W_1^\textrm{crit}=0.045$,
leads to a closed loop of immiscibility. The nematic phase rich in
small platelets (N$_2$) and the paranematic phase have merged.

\begin{figure}
  \centering 
  \subfigure{\includegraphics[width=0.22\textwidth,clip]{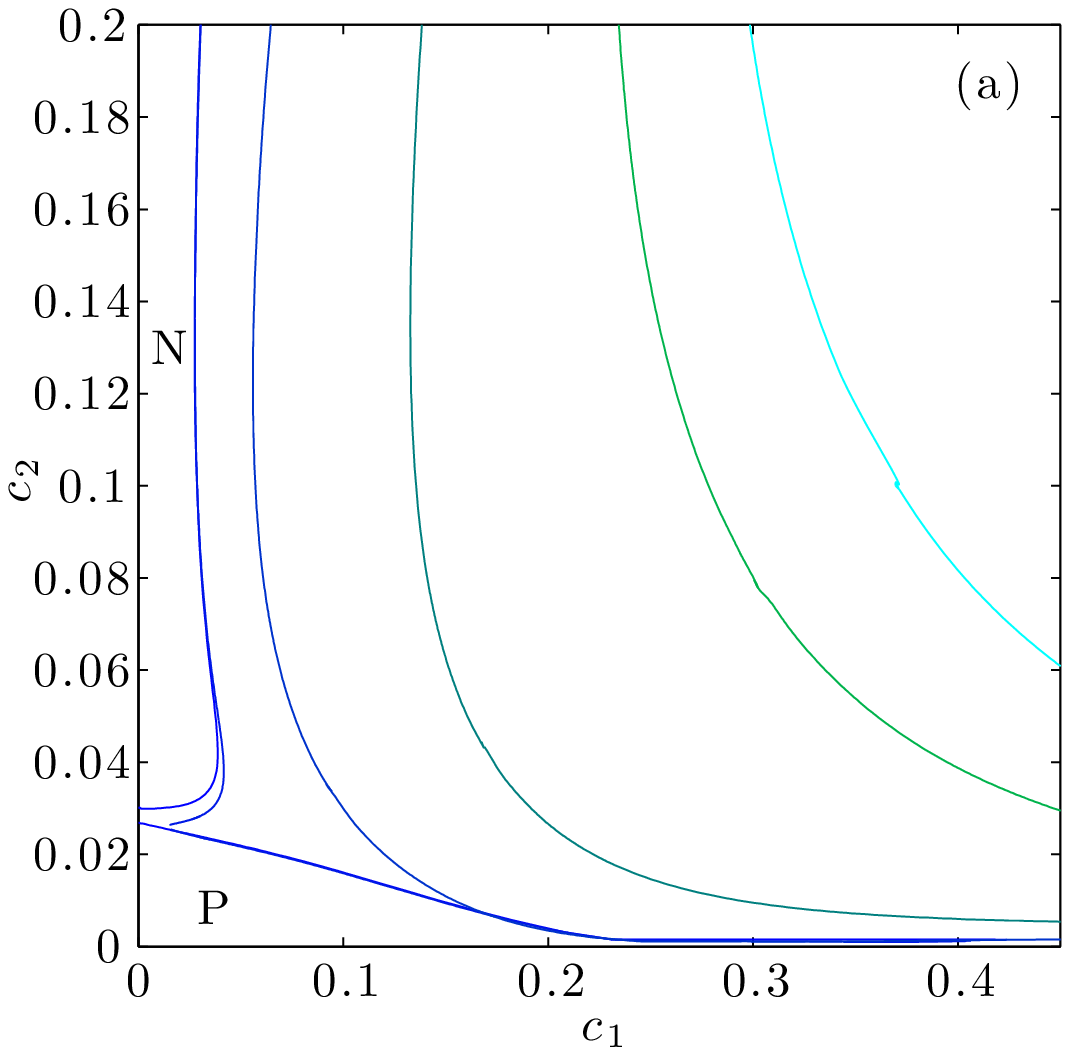}}
  \subfigure{\includegraphics[width=0.22\textwidth,clip]{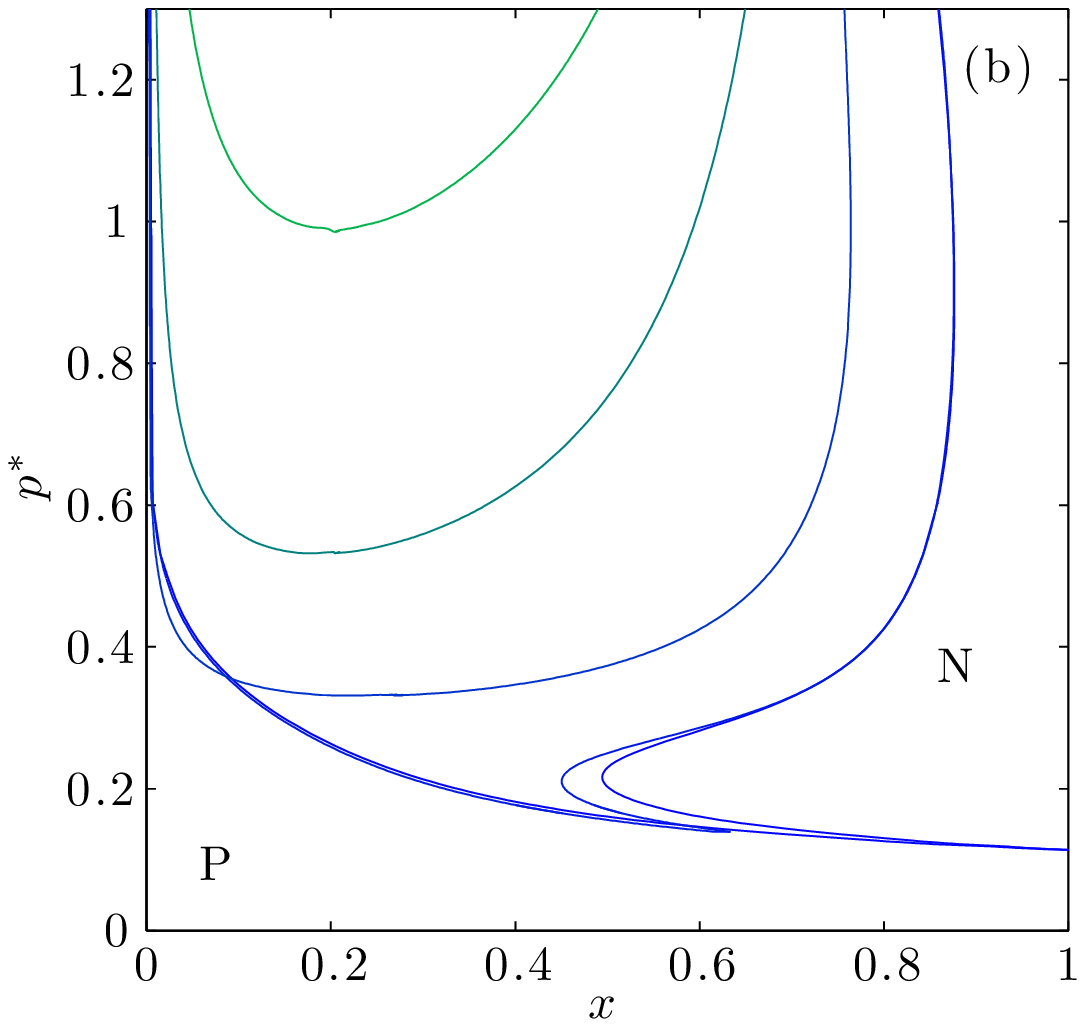}}
  \subfigure{\includegraphics[width=0.22\textwidth,clip]{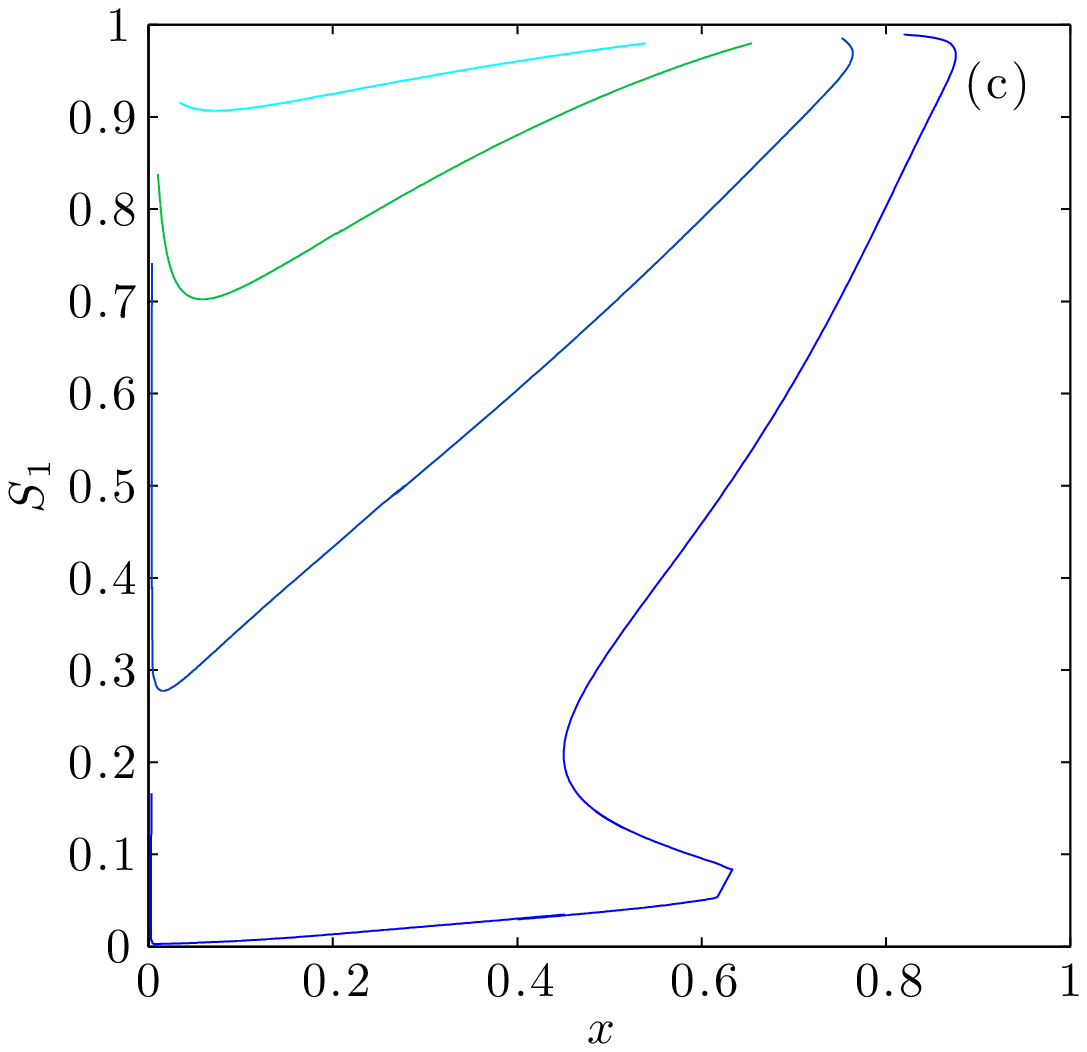}}
  \subfigure{\includegraphics[width=0.22\textwidth,clip]{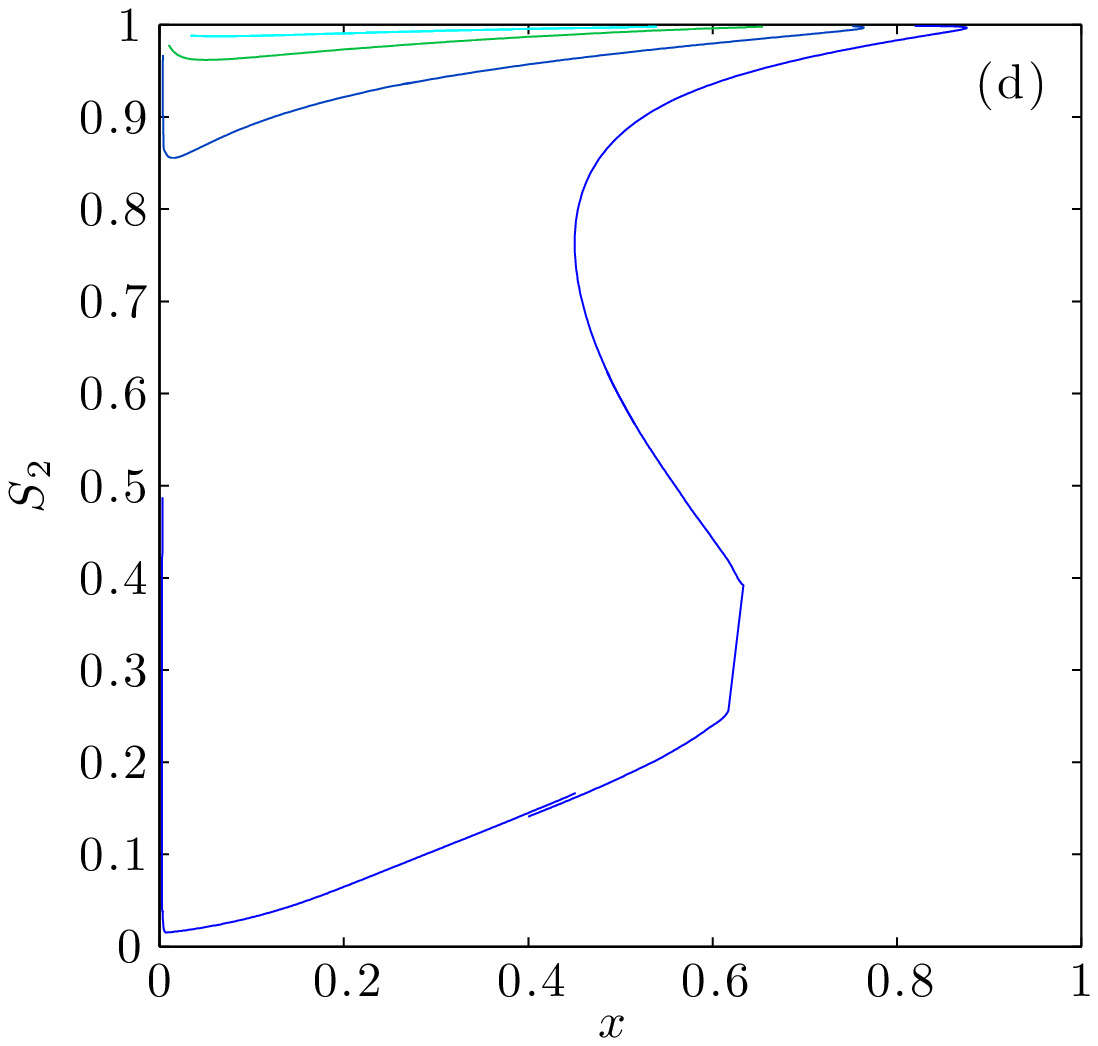}}
  \caption{Same as Fig.~\ref{figure2} except for $\lambda=2.5$ and for
    external field strengths $W_1=0,0.01,1,2.5,5$ and $7$. In (c) and
    (d) we only present partial order parameters for $W_1=0.01,1,2.5$
    and $5$ for clarity. The small kinks at low values of $S_i$ in (c)
    and (d) are due to numerical artifacts.}
\label{figure4} 
\end{figure}

In Fig.~\ref{figure4} we present results for $\lambda=2.5$. The
fieldless case displays I-N coexistence, an I-N-N triple point and
coexistence between two nematic states, which does not end in a
critical point. Applying just a small field $W_1=0.01$ has a
considerable effect on the phase behaviour: the pure system of species
2 looses the P-N transition as $W_1=0.01 >
W_1^\textrm{crit}=0.0069$. However, the transition persists in the
pure system of species 1. Hence, there is still P-N coexistence and
indeed a tail between about $x=0.5$ and $x=0.6$ where reentrant
behaviour occurs. For $W_1=1$, there is no P-N transition for either
of the pure limits. There is, however, a large immiscibitity gap
between two distinct nematic phases, $N_1$ and $N_2$. Increasing the
field strength raises the phase coexistence to higher pressures and
narrows the phase coexistence region. However, large steps in field
strength are required to have a significant effect on the
system. $W_1=7$ (which corresponds to $W_2=49$) is approximately 45
times stronger than the field reqired to homogenise the system at
$\lambda = 1.5$ and yet even with such a high field strength, there
persists a wide coexistence region. The order parameter curves
approach very high values (close to 1) for large values of $W_1$. This
forms a limit to the densities at which we may probe at $\lambda=2.5$:
for very high order parameters it is numerically difficult to
determine the ODFs, even with very fine $\theta$-grids.

\section{Conclusions and Outlook}
\label{sec:conclusions}
We have investigated the effects of an external aligning (magnetic)
field on the phase behaviour of binary mixtures of circular hard core
platelets with zero thickness. Using the FMT of
Refs.~\cite{esztermann06,phillips.j:2010.a}, we have traced
paranematic and nematic phase boundaries and have examined the partial
nematic order parameters at coexistence. Three different
representative values for the radial bidispersity
($\lambda=1.5,2,2.5$) have been studied. The topologies for each of
these values are different from each other in the fieldless case. For
the smallest size ratio considered, $\lambda=1.5$, the fieldless case
shows only I-N phase coexistence over the entire range of
compositions. For $\lambda=2$, besides I-N coexistence there is also
N-N coexistence ending in an upper critical point and an I-N-N triple
point occurs. For $\lambda=2.5$ the N-N phase coexistence does no
longer end in a critical point (at least up to the densities we
considered). Applying the external field induces paranematic order in
the low-density regions of the phase diagrams (which are isotropic in
the absence of a field). Increasing the field strength leads in all
cases to a narrowing of the biphasic P-N region. The P-N transition
further destabilises upon increasing the external field strength.  The
system becomes more strongly ordered, such that for $\lambda=1.5$ and
2 phase coexistence disappears (at a double critical point) and the
system is in a single-phase nematic state for all statepoints. The
field strength required to complete the homogenisation increases for
increasing size ratio. In contrast, for $\lambda=2.5$, the coexisting
nematic states become so well-ordered that the system does not become
homogeneously nematic up to the field strengths we have applied.
 
Results from computer simulation studies for this (or a similar) model
mixture are highly desirable, as are experimental studies. Colloidal
platelets are often significantly polydisperse in both radius and
thickness, see for example Ref.~\cite{vanderkooij.f.m:2000.a}, so
effects due to polydispersity will play a role in experimental
systems, which are not accounted for in the present theory. Recently,
the P-N interface in suspensions of boardlike goethite particles has
been investigated experimentally \cite{vandenpol.e:prep}. Anticipating
that similar studies could be made in systems of colloidal platelets,
a further exciting avenue would be to investigate the properties of
the P-N interface using FMT, which has already been shown to compare
well with experimental and simulation results for the I-N interface
\cite{vanderbeek.d:2006.b}.  It would be interesting to consider in
theoretical work the effects that are induced by finite thickness of
the platelets. In the present study we restricted ourselves to
mixtures with moderate size asymmetry, as we expect the theory to
describe these accurately. Investigating highly asymmetric mixtures,
possibly based on the depletion picture, is an interesting issue for
future work.

\begin{appendix}
\section{Self-Consistency Equations for the Orientational Distribution Functions}
We give a summary of the equations that are necessary to find the ODFs
at a given composition, $x$ and concetration, $c$. The excess free
energy from FMT is the sum of the right hand sides of Eqs.~(13) and
(27) of Ref.~\cite{phillips.j:2010.a}. Inserting this, together with
the ideal free energy (\ref{eq:ideal}) and the external potential
(\ref{eq:external_potential}), into the grand potential functional
(\ref{eq:grand_potential}), and employing the minimisation principle
(\ref{eq:minimization_principle}) leads to two coupled Euler-Lagrange
equations for the ODFs:
\begin{widetext}
\begin{align}
\label{eq:self_consistency_1}
\Psi_{1}(\theta) & = \frac{1}{Z_{1}} \exp \bigg[  -8\pi c \int_{0}^{\pi/2} d \theta '  \sin \theta '  K(\theta, \theta ')[(1-x) \Psi_{1}(\theta ') + \frac{1}{2}x(\lambda^{2} + \lambda)\Psi_{2}(\theta ')  ] \notag \\ &  -32\pi c^{2} \int_{0}^{\pi/2} d \theta '  \sin \theta ' \int_{0}^{\pi/2} d \theta ''  \sin \theta ''   L(\theta, \theta ', \theta'') \notag \\ &  [(1-x)^{2}\Psi_{1}(\theta')\Psi_{1}(\theta'')+2x(1-x)\lambda^{2}\Psi_{1}(\theta')\Psi_{2}(\theta'')+x^{2}\lambda^{4}\Psi_{2}(\theta')\Psi_{2}(\theta'') ] +W_1\sin^{2}\theta \bigg],
\end{align}
\begin{align}
\label{eq:self_consistency_2}
 \Psi_{2}(\theta) & = \frac{1}{Z_{2}} \exp \bigg[ -8 \pi c \int_{0}^{\pi/2} d \theta '  \sin \theta ' K(\theta, \theta ')[x \lambda^{3}\Psi_{2}(\theta ') + \frac{1}{2}(1-x) (\lambda^{2} + \lambda)\Psi_{1}(\theta ') \notag \\ & -32\pi c^{2} \int_{0}^{\pi/2} d \theta '  \sin \theta ' \int_{0}^{\pi/2} d \theta ''  \sin \theta '' L(\theta, \theta ', \theta'') \notag \\ &  [x^{2} \lambda^{6}\Psi_{2}(\theta')\Psi_{2}(\theta'')+2x(1-x)\lambda^{4}\Psi_{1}(\theta')\Psi_{2}(\theta'')+(1-x)^{2}\lambda^{2}\Psi_{1}(\theta')\Psi_{1}(\theta'') ] +W_2\sin^{2}\theta \bigg],
\end{align}
\end{widetext}
where the constants $Z_{1}$ and $Z_{2}$ are such that the
normalisation $\int d
\boldsymbol{\omega}\Psi_{i}(\boldsymbol{\omega})=1$, for $i=1,2$. The
numerical procedure is the same as that described in
Ref.~\cite{phillips.j:2010.a}, which is an extension of the procedure
introduced in Ref.~\cite{herzfeld84}.  The integral kernel
$K(\theta,\theta ')$ is
\begin{align}
\label{eq:kernel_k}
K(\theta,\theta')=&\int_{0}^{2\pi}d\phi \sin\gamma=\int_{0}^{2\pi}d\phi \sqrt{1- (\boldsymbol{\omega} \cdot \boldsymbol{\omega}')^{2}}
\notag \\ & =\int_{0}^{2\pi}d\phi \sqrt{1-(\cos\theta \cos\theta' + \sin\theta \sin\theta' \cos\phi)^{2}},
\end{align}
where $\phi$ is the difference between the azimuthal angles of the two
platelets and the kernel $L(\theta,\theta', \theta'')$ is
\begin{widetext}
\begin{align}
\label{eq:kernel_l}
L(\theta,\theta', \theta'')=&\int_{0}^{2\pi} \int_{0}^{2\pi} d\phi' d\phi''  | \boldsymbol{\omega} \cdot ( \boldsymbol{\omega}' \times \boldsymbol{\omega}'')| \notag \\ &\hspace{-4.2mm}=\int_{0}^{2\pi} \int_{0}^{2\pi} d\phi' d\phi''  |\sin \theta(\sin \phi' \sin \theta' \cos \theta''  \notag \\ &\hspace{5mm}+\cos\theta' \sin \phi'' \sin \theta'' )+\cos\theta(\cos\phi' \sin\theta' \sin\phi '' \sin \theta '' \notag \\ &\hspace{5mm}- \sin \phi' \sin \theta' \cos \phi '' \sin \theta '')|.
\end{align}
\end{widetext}
The solutions of Eqs.~(\ref{eq:self_consistency_1}) and (\ref{eq:self_consistency_2}), $\Psi_{1}(\theta)$ and $\Psi_{2}(\theta)$, are then inserted into the equations $p^{\textit{A}}=p^{\textit{B}}$ and $\mu_i^{\textit{A}}=\mu_i^{\textit{B}}$, $i=1,2$ in order to find the coexisting states. Triple points are located where the P-N binodals and the N-N binodals intersect.
\end{appendix}

\begin{acknowledgments}
  It is a great pleasure for us to dedicate this paper to Professor
  Henk Lekkerkerker on the occasion of his 65th birthday. Both Henk's
  scientific work and his personality provided much inspiration and
  motivation for carrying out the present study as well as related
  work by the authors. We also thank Nigel Wilding, Mark Miller, Chris
  Newton and Susanne Klein and the Liquid Crystals group of HP Labs,
  Bristol for useful discussions. Financial support from the EPSRC,
  from HP Labs, Bristol and via the SFB840/A3 of the DFG is gratefully
  acknowledged.
\end{acknowledgments}

\end{document}